\documentclass[sigconf]{acmart}

\AtBeginDocument{%
  \providecommand\BibTeX{{%
    \normalfont B\kern-0.5em{\scshape i\kern-0.25em b}\kern-0.8em\TeX}}}

\copyrightyear{2022}
\acmYear{2022}
\setcopyright{rightsretained}
\acmConference[ASSETS '22]{The 24th International ACM SIGACCESS Conference on Computers and Accessibility}{October 23--26, 2022}{Athens, Greece}
\acmBooktitle{The 24th International ACM SIGACCESS Conference on Computers and Accessibility (ASSETS '22), October 23--26, 2022, Athens, Greece}\acmDOI{10.1145/3517428.3550398}
\acmISBN{978-1-4503-9258-7/22/10}

\usepackage{graphics}
\usepackage{enumitem}
\usepackage{balance}
\usepackage{xspace}
\usepackage{color, soul}
\usepackage{fixltx2e}

\def\ie{\textit{i.e.,}\xspace}
\def\etal{\textit{et al.}\xspace}

\def\eg{\textit{e.g.,}\xspace}

\begin{document}

\begin{teaserfigure}
\centering
    \vspace{-0.1in}
    \includegraphics[
    width=.9\textwidth]{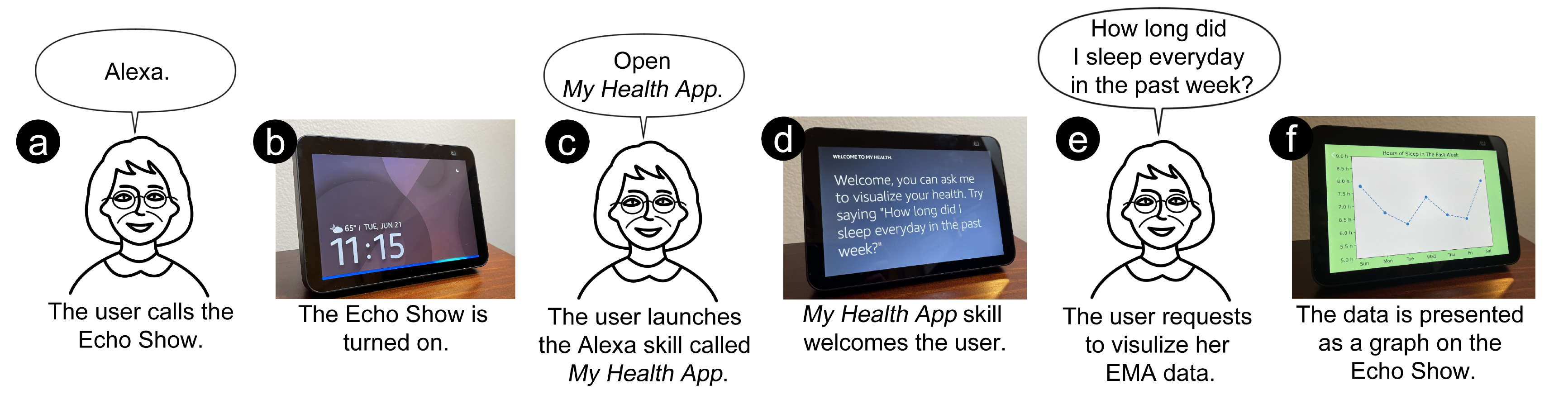}
    \vspace{-0.2in}
    \caption{With touchscreen based standalone voice--first virtual assistant, older adults are able to query and visualize the time--series based Ecological Momentary Assessment (EMA) data (\eg~the quality and time of the sleep).}
    \label{fig::teaser}
\end{teaserfigure}

\title{Towards Visualization of Time--Series \\ Ecological Momentary Assessment (EMA) Data on Standalone Voice--First Virtual Assistants}

\author{Yichen Han$^{1}$, Christopher Bo Han$^{3}$, Chen Chen$^{1}$, Peng Wei Lee$^{2}$, \\Michael Hogarth$^{5}$, Alison A. Moore$^{5}$, Nadir Weibel$^{1}$, Emilia Farcas$^{4}$}

\affiliation{%
\vspace{0.2cm}
  \department{$^{1}$Computer Science and Engineering, $^{2}$Electrical and Computer Engineering, \\$^{3}$Department of Mathematics, $^{4}$Qualcomm Institute, $^{5}$School of Medicine}
  \city{University of California San Diego, La Jolla}
  \state{California}
  \country{United States}
}
\email{{y4han, cbhan, chenchen, pwlee, mihogarth, alm123, efarcas, weibel}@ucsd.edu}

\renewcommand{\shortauthors}{Han \etal}

\begin{abstract}

Population aging is an increasingly important consideration for health care in the 21th century, and continuing to have access and interact with digital health information is a key challenge for aging populations. 
Voice-based Intelligent Virtual Assistants (IVAs) are promising to improve the Quality of Life (QoL) of older adults, and coupled with Ecological Momentary Assessments (EMA) they can be effective to collect important health information from older adults, especially when it comes to repeated time-based events. 
However, this same EMA data is hard to access for the older adult: although the newest IVAs are equipped with a display, the effectiveness of visualizing time--series based EMA data on standalone IVAs has not been explored. 
To investigate the potential opportunities for visualizing time--series based EMA data on standalone IVAs, we designed a prototype system, where older adults are able to query and examine the time--series EMA data on Amazon Echo Show --- a widely used commercially available standalone screen--based IVA.
We conducted a preliminary semi--structured interview with a geriatrician and an older adult, and identified three findings that should be carefully considered when designing such visualizations.
\end{abstract}


\maketitle

\balance
\begin{figure*}[t]
    \centering
    \includegraphics[width=\textwidth]{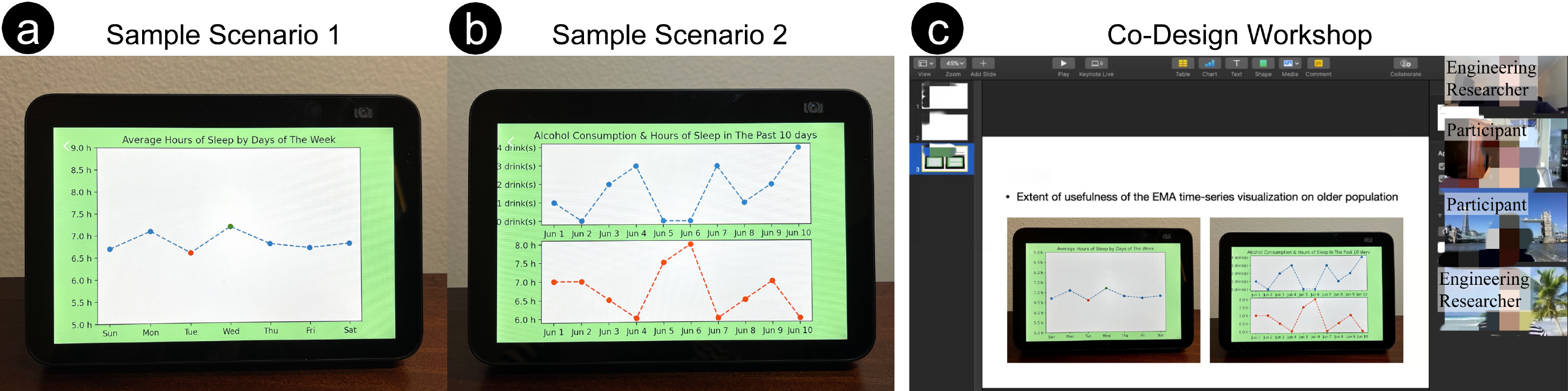}
    \vspace{-0.2in}
    \caption{Two sample demonstrations on visualizing time-series data on Amazon Echo Show with voice input (a-b) and a semi-structured interview exploring their utility (c).}
    \label{fig::scenarios}
\end{figure*}

\section{Introduction}
Population aging is a global health consideration in the 21th century~\cite{dalgaard2022physiological}, especially when it comes to keeping track of older adults' daily health data. 
To more effectively support older adults and their clinicians to collect daily health check-ins, studies have shown the effectiveness of using Ecological Momentary Assessments~(EMA), a well--known method to more conveniently assess specific recurring behaviors and track specific health states. 
An effective EMA strategy could allow clinicians to constantly monitor older adults and ensure their physical and mental well-being, hopefully leading to an increase in their Quality of Life~(QoL)~\cite{evangelista2015examining, salzmann2012panoptic}. 

However, such strategies often necessitates tremendous amounts of time and effort; and further issues, such as communication challenges, could arise~\cite{Sundler2016}.
While a wide variety of technologies have been used to keep track of the older adults' conditions~\cite{mortenson2015power, sixsmith2013technologies}, many older adults are still facing obstacles while using smart devices, most notably those with complicated Graphical User Interfaces (GUI)~\cite{leung2012older, li2020older}.

Voice based conversational user interfaces offers an promising alternative interaction channel for older adults to interact with their digital health information.
Such modality allows older adults to naturally and easily use voice based conversation--like commands to delegate tasks to, or query information from their healthcare providers.
While Trajkova~\etal~\cite{Milka2020} demonstrates the limitations of today's voice assistants among aging users, others~(\eg~\cite{Chen2021, Charles2021, Lifset2020, Mrini2021, adaimi2020usability})~show the feasibility and potential usefulness of using standalone voice--based Intelligent Virtual Assistants~(IVAs) for conducting EMA among older adults patients.

Despite the potential of deploying EMAs through IVAs, due to the nature of ambiguity of conversational voice user interfaces, Chen~\etal~\cite{Chen2021, Chen2021toward} pointed out the necessity of also integrating simple visual components for facilitating better interaction with the EMAs deployed on the device. 
The same group of researchers~\cite{Chen2021toward} also investigated how to visualize simple interaction elements, such as the texts of EMA questions and input buttons on Amazon Echo Show~\cite{echo_show}, a widely used commercially available standalone IVAs.
However, while data collection through EMA is important, previous work did not yet explore how to designs and visualize the data collected through EMAs on the IVA device itself, especially when it comes to display time--series visualizations.

In this paper we present the {\it first} preliminary exploratory effort to design time--series data visualizations on screen based voice--first IVAs.
We chose Amazon Echo Show~\cite{echo_show} as the standalone voice--based IVA device with a built--in touchscreen due to its dominant market share~\cite{alexa_marketshare}, however, many of our findings could be transferred to other screen--based voice--first IVAs.
We prototyped a system that enables older adults to examine their past EMA data on the built--in touchscreen of Echo Show.
By observing interaction with the IVA visualization, and especially older adults' past data and corresponding trends, we found that older adults might be able to make {\it in-situ} decisions to address possible unhealthy lifestyle.

We conducted a remote co--design workshop with two representative stakeholders -- an older adult and a geriatrician.
After thematic analysis of the recordings, we present findings and potential future development. 
\begin{figure*}[t]
    \centering
    \includegraphics[width=0.7\textwidth]{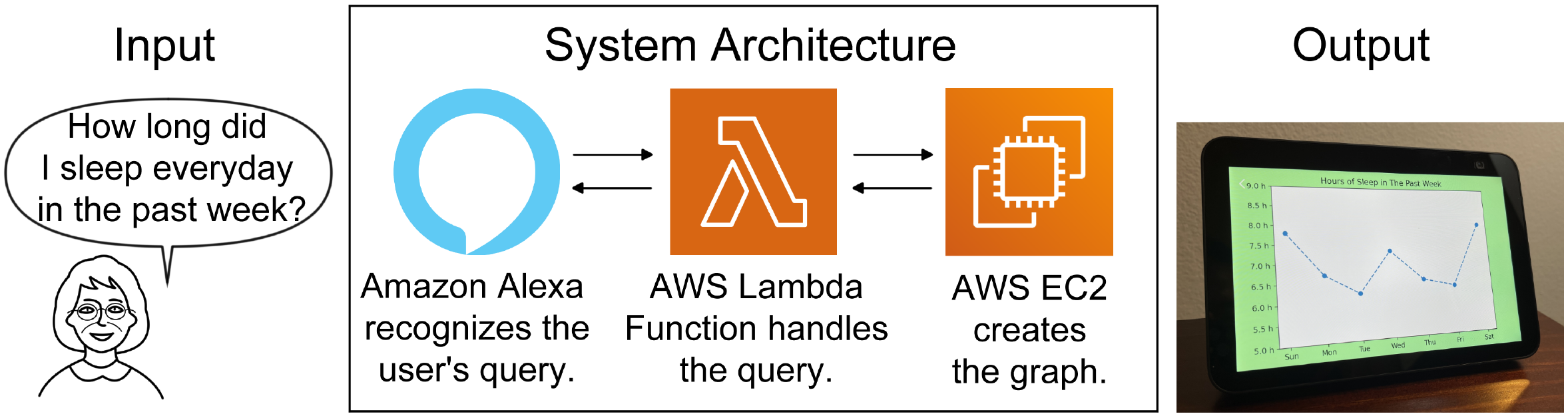}
    \vspace{-0.1in}
    \caption{When older adults request a graph from voice assistant Alexa. AWS Lambda handles this request by asking EC2 to generate a graph and send it back to Echo show, paired with Alexa. Faces are pixelized to protect participants' personally identifiable information.}
    \vspace{-0.1in}
    \label{fig::architecture}
\end{figure*}

\section{Preliminary Design}

Our preliminary designs consist of two example scenarios demonstrating how time--series EMA data can be visualized and are beneficial to older adults (see Figs.~\ref{fig::scenarios}a and b). 
In both cases, we assume that the EMA data has already been collected and stored in a secure Electronic Health Record~(EHR). 
We considered scenarios of {\bf one--measure visualization}, where we only visualize one measure (\eg~ the hour of sleep as shown in Fig.~\ref{fig::scenarios}a) over the time domain, and {\bf two--measure visualization}, that places two one--measure graphs with the same dimension vertically, which might indicate possible correlations between two dependent variables. 
Visualizations with more than two measures are out of our scope due to the consideration of the screen size and participants' cognitive load.

\vspace{+0.05in}
\noindent{\bf Scenario 1 -- One--Measure Visualization:}
Fig.~\ref{fig::scenarios}a demonstrates a one--measure visualization for hours of sleep, one important time-series EMA data used in geriatrics to help tracking older adults' QoL~\cite{Crowley2011}. 
A voice skill could be initiated by a simple query: {\it``Hey Alexa, what's my average hours of sleep everyday in the week?''}
The mean hours of sleep by days in a week will then be computed, yielding a line plot shown on the IVA's screen.
The trend of the graph could help older adults recall the common activities on particular days in a week that possibly led to an irregularity of sleeping hours.

\vspace{+0.05in}
\noindent{\bf Scenario 2 -- Two--Measure Visualization:}
Older adults should also be able to visualize two--measure plots with the same time span to examine any potential correlations. 
As an example, Fig.~\ref{fig::scenarios}b demonstrates how older adults could query about their sleep hours and alcohol consumption in the past $10$ days. 
This plot can visually guide the user to find the correlation between the daily alcohol consumption and sleep, so that the user can properly {\it in--situ} adjust to a healthier lifestyle. 

\vspace{+0.05in}
\noindent{\bf Prototype:}
We prototyped the system based on Amazon Alexa and Amazon Web Services (AWS). 
The overall system architecture is shown in Fig.~\ref{fig::architecture}, and consists of three major components: (1) a front--end converting users' utterances to queries, (2) a middle--tier that handles the queries and requests the graphs, and (3) a back--end to generate the graphs. 
In the front--end, speech recognition technology in the IVAs, such as Alexa Voice Service (AVS) in an Echo Show, are used to recognize the raw audio input. 
An AWS Lambda Function is then triggered and passes through the transcribed text to the back-end server via RESTful API which will then generate a time-series graph. 
Through Amazon Presentation Language (APL)~\cite{apl}, the plot is then rendered on the touchscreen as a background image.
Through this process, older adults are able to view their time--series healthcare information as a graph on their standalone devices. 
\section{Design and User Evaluation}

To understand how data visualization can help the aging population, we conducted a remote 30--minute co--design workshop through Zoom\footnote{Zoom: \url{https://zoom.us}.}~\cite{Yarmand2021}, and interviewed an older adult~(P1) from \href{https://www.viliving.com/locations/ca/san-diego-la-jolla}{the retirement community "Vi" in La Jolla, California} and a geriatric provider~(P2) from \href{https://health.ucsd.edu/Pages/default.aspx}{the University of California San Diego Health System}. Our study has been approved by the Institutional Review Board~(IRB).
We showed the interviewees our two initial visualization prototypes (see Fig.~\ref{fig::scenarios}a and b) and solicited feedback, encouraging them to propose alternatives.
Specifically, the participants were asked to express their opinions regarding \textbf{(a)} the potential functionality and effects of the proposed EMA visualizations, as well as \textbf{(b)} how useful these visualizations would be to their daily routine.
Fig.~\ref{fig::scenarios}c shows a screenshot of our Zoom session with the two stakeholders. 
\section{Preliminary Findings}

We recorded the complete Zoom session and analyzed the discussion using analysis~\cite{Joffe2012}. In this section we summarize our findings in terms of three aspects.

\vspace{+0.05in}
\noindent{\bf (1) Selections of Measures to be Visualized Should be Carefully Considered. }\\
Participants agreed that mean values, as visualized in Fig.~\ref{fig::scenarios}a, could provide clear insights while retrospecting the older adults' activities over the previous weeks. 
However, participants also expressed a preference for seeing additional measures to help evaluating the spread of the measured data.
For example, P2 mentioned that if {\it``error bars''} existed in Fig.~\ref{fig::scenarios}a,{\it``the  preciseness of the number of sleeping hours''} could be shown.
This revision would help visualize the consistency of sleeping time among aging populations.

\vspace{+0.5in}
\noindent{\bf(2) The Span of Horizontal and Vertical Axis Should be Carefully Designed.}\\
Participants suggested to have more flexibility in terms of the range for the vertical axis. 
As shown in Fig.~\ref{fig::scenarios}a, the range of the vertical axis is a four--hour difference. 
Changing the range of the vertical axis to be more flexible would allow users to display a more accurate variability of the number of sleeping hours.
P2 mentioned that having a set vertical span has the limitation of being {\it``bounded by the responses they [the users] have''} and emphasized that having a flexible vertical range could be implemented {\it``so people don't potentially freak out''}.

Participants also emphasized the importance of carefully designing the time span (\ie~visualized as horizontal axis).
In Figs.~\ref{fig::scenarios}a and b, the graphs show data within only a short time span (\eg~the time span of Fig.~\ref{fig::scenarios}a is a week and the time span of Fig.~\ref{fig::scenarios}b is 10 days).
However, healthcare providers could request data from a longer time span, such as from the past several months or years. 
P2 in fact questioned the effectiveness of graphing only tens days of data in Fig.~\ref{fig::scenarios}b and suggested to {\it``make the axis longer so you[the visualization] show more''} health data. Future designs should adapt the time span according to the user's demand.

\vspace{+0.05in}
\noindent{\bf(3) The Types of Data to be Visualized Should be Carefully Chosen.}\\
Currently, the prototypes only display quantitative data, but there are numerous essential qualitative data points and types that could be included in the graphs, as suggested by the participants.
The quality of sleep, an example of qualitative data, can provide a clearer picture on how alcohol is affecting the user's sleep.
This data type, however, is difficult to be implemented in the visualizations. One method is to {\it``build it in as a color ... to overlay the number of [sleeping] hours with quality''}, as suggested by P2.
The addition of "setting goals" perhaps overlaid on top of the quantitative data would also make the visualizations more personal to the user, since meeting their goal encourages the user to maintain a better lifestyle. P1 mentioned that {\it ``you might set your goal fairly low... so if you're consistently meeting your lower goal, then that gives you a chance to consider whether you want to increase the goal or not''}. 
In the graphs showing the correlation between the amount of consumed alcohol and the sleeping time, the time difference between when the user last consumed alcohol and the time they fell asleep the same day is also a factor that affects sleep. Therefore, this variable could also be considered to add into the visualizations.
\section{Conclusion and Future Work}
To help older adults visualize their time-series EMA data through voice input, we designed and implemented a prototype that can generate one--measure and two--measure graphs on standalone voice--first virtual assistants. 
We explored opportunities of the system and summarized needs on 3 major components of this visualization, which are measures, types, and time span of data. 
Although some of these opportunities can be easily implemented, others like providing a better picture of the user's conditions and goals, can be difficult to integrate into the visualizations and interpreted correctly by the user and the healthcare providers. 

This work is only scratching the surface in terms of how to design graphical visualization of EMAs on IVAs, and outline a number of challenges that still need to be resolved. 
In future work, we plan to conduct user studies with older participants to fully explore opportunities and challenges faced by this a system like the one we have prototyped.
Meanwhile, integrating some straight-forward suggested refinements as part of our existing prototype will already make visualizations more adaptive to context and representation of qualitative data. 
\begin{acks}
This work is part of project \href{http://voli.ucsd.edu}{VOLI}~\cite{voli_website} and was supported by NIH/NIA under grant R56AG067393.
Co-author Michael Hogarth has an equity interest in LifeLink Inc. and also serves on the company’s Scientific Advisory Board. The terms of this arrangement have been reviewed and approved by the UC San Diego in accordance with its conflict of interest policies.
We appreciate insightful feedback from the anonymous reviewers, Manas Satish Bedmutha, Mary Draper and fellow colleagues from \href{http://designlab.ucsd.edu}{the Design Lab} and \href{https://cse.ucsd.edu}{Computer Science and Engineering} at \href{https://ucsd.edu}{UC San Diego}, as well as residents from \mbox{the Vi in La Jolla}.
\end{acks}

\bibliographystyle{ACM-Reference-Format}
\bibliography{reference}

\end{document}